\documentclass[preprintnumbers,nofootinbib,superscriptaddress,10.5pt]{revtex4}
\usepackage[dvipdfmx]{graphicx} 
\usepackage{amsmath,amssymb,amsfonts, bm}

  \def\g{\gamma}  \def\d{\delta}  \def\e{\epsilon}   \def\th{\theta}     \def\m{\mu} \def\n{\nu}     \def\r{\rho}   \def\t{\tau}       

\def\dg{\dagger}

    \newcommand{\To}{\Rightarrow}

\def\abs#1{\left| #1\right|}

\renewcommand{\Im}{{\rm Im}\,}

\begin{document}
%%%%%%%%%%%%%%%%%%%%%%%%%%%%%%%%%%%%%%%%

\title{
A Rephasing Invariant Formula for the Dirac CP Phase and \\ General Perturbative Expansion: Prospects for DUNE and T2HK}

\preprint{STUPP-25-282}
%%%%%%%%%%%%%%%%%%%%%%%%%%%%%%%%%%%%%%%%

\author{Masaki J. S. Yang}
\email{mjsyang@mail.saitama-u.ac.jp}
\affiliation{Department of Physics, Saitama University, 
Shimo-okubo, Sakura-ku, Saitama, 338-8570, Japan}
\affiliation{Department of Physics, Graduate School of Engineering Science,
Yokohama National University, Yokohama, 240-8501, Japan}

%%%%%%%%%%%%%%%%%%%%%%%%%%%%%%%%%%%%%%%%

%\date{\today}

%%%%%%%%%%%%%%%%%%%%%%%%%%%%%
\begin{abstract} %%%%%%%%%%%%%%%%%%%%%
%%%%%%%%%%%%%%%%%%%%%%%%%%%%%

We present a formula for the Dirac CP phase 
$\delta  = \arg ( U_{e1} U_{e2} U_{\mu 3} U_{\tau 3} / U_{e3} \det U_{\rm MNS} )$, 
directly derived from the lepton mixing matrix in an arbitrary basis of phases. 
In contrast to the numerically suppressed Jarlskog invariant, this expression is computationally simple and less sensitive to approximations. 
We apply the formula to derive general perturbative corrections from charged-lepton mixing $s_{ij}^{e}$ to the underlying CP phase of neutrinos $\delta_{\nu}$. 
A compact analytic expressions $\delta = \delta_{\nu} + s_{12}^{e} D_{12} + s_{13}^{e} D_{13} + s_{23}^{e} D_{23}$ shows that these corrections can substantially exceed $O(10^{\circ})$, potentially within the reach of future long-baseline experiments such as DUNE and T2HK.

%%%%%%%%%%%%%%%%%%%%%%%%%%%%%
\end{abstract} %%%%%%%%%%%%%%%%%%%%%%
%%%%%%%%%%%%%%%%%%%%%%%%%%%%%

\maketitle

%%%%%%%%%%%%%%%
\section{Introduction}
%%%%%%%%%%%%%%%

The phenomenon of CP violation has been one of the cornerstones in the progress of particle physics \cite{Kobayashi:1973fv}. 
Current long-baseline experiments such as NO$\nu$A \cite{NOvA:2021nfi} and T2K \cite{T2K:2021xwb} are measuring the Dirac CP phase $\delta$ in the lepton sector. However, recent  results show a significant tension under the assumption of the normal mass hierarchy \cite{Esteban:2024eli}.
The upcoming Deep Underground Neutrino Experiment (DUNE) \cite{DUNE:2020jqi} and Tokai to Hyper-Kamiokande (T2HK) \cite{Hyper-KamiokandeProto-:2015xww} 
are expected to have 
a discovery potential for $|\sin \delta| \gtrsim 0.5$ and a resolution of $O (10^\circ)$ with ten years of data.
In the next decade, the phase is one of the most promising physical quantities to be observed.

Although many studies have explored the behavior of the CP phase of mixing matrices \cite{Ge:2011qn, Feruglio:2012cw, Holthausen:2012dk, Denton:2020igp},  a general analytic understanding has remained elusive due to the large number of free parameters. 
Various studies have evaluated perturbations by charged leptons for a range of models and characteristic mixing matrices \cite{Xing:2002sw, Antusch:2005kw, Farzan:2006vj, Hochmuth:2007wq, Marzocca:2013cr,  Petcov:2014laa, Dasgupta:2014ula}. 
However, most approaches rely on the Jarlskog invariant \cite{Jarlskog:1985ht}, 
which is as small as $O(10^{-5})$ and $O(10^{-2})$ in the quark and lepton sector.
Since the invariant is particularly sensitive to various approximations, 
it requires careful treatment to maintain computational accuracy. 
Moreover, since this invariant does not contain information on the sign of $\cos\delta$, 
this sign must be calculated separately to fully reproduce the experimental results.

In this letter, we derive a formula that represents the Dirac phase $\delta$ 
by arguments of the mixing matrix elements. 
This formula simplifies the calculation of the phase and reduces the impact of numerical errors. 
As a specific application, a general perturbative expansion in mixing parameters of charged leptons $s_{ij}^e$ shows that they can induce significant corrections to the Dirac phase probed in upcoming experiments.

%%%%%%%%%%%%%%%%%%%&&&&&&&&&
\section{A Formula and Its Perturbative Expansion}
%%%%%%%%%%%%%%%%%%%&&&&&&&&&

We first present a method to extract the Dirac CP phase $\delta$ directly from the lepton mixing matrix $U_{\rm MNS}$ in an arbitrary basis of phases. 
To convert  $U_{\rm MNS}$ into its PDG parametrization $U_{\rm MNS}^{0}$, 
general phase transformations remove unphysical phases as
\begin{align}
U_{\rm MNS}^{0} = \Psi_{L}^{\dagger} U_{\rm MNS} \Psi_{R} \, , 
\end{align}
where $\Psi_{L,R} = {\rm diag} (e^{i \g_{(L, R)1}}, \, e^{i \g_{(L,R)2}} , \, e^{i \g_{(L,R)3}})$
 with phases $\gamma_{(L,R)i}$.
By the redefinition of an overall phase $\g_{(L,R)i} \to \g_{(L,R)i} + \g$, the number of independent degrees of freedom is five.
In the standard PDG parametrization, the matrix elements satisfy the following conditions.
\begin{align}
\arg U_{e1}^{0} = \arg U_{e2}^{0} = \arg U_{\m 3}^{0} = \arg U_{\t 3}^{0} = 0  \, , ~~
\arg [ U_{\m 1}^{0} U_{\t 2}^{0} - U_{\m 2}^{0} U_{\t 1}^{0} ] = \arg [U_{e3}^{0 *} \det U_{\rm MNS}^{0}] = \d \, . 
\end{align}
The last condition holds as a consequence of the matrix inversion formula. 

By fixing the five phases $\gamma_{(L,R)i}$ from these five conditions, 
the Dirac CP phase in an arbitrary basis of phases is expressed as follows, 
\begin{align}
\d & = \arg  \left [  { U_{e1} U_{e2} U_{\m 3} U_{\t 3} \over U_{e3} \det U_{\rm MNS} } \right ]  \, .\label{d2}
%& = \arg U_{e1} + \arg U_{e2} + \arg  U_{\m 3} + \arg  U_{\t 3}  - \arg U_{e3}  - \arg (\det U_{\rm MNS}) \, , 
\end{align}
This expression is manifestly rephasing invariant \cite{Wu:1985ea,Bjorken:1987tr, Nieves:1987pp, Jenkins:2007ip} including the phase of $\det U_{\rm MNS}$,  
 and reduces to the Dirac phase $\delta$ in the standard PDG convention. 
When the complex conjugate of the denominator is multiplied by the numerator, 
this is an eighth-order rephasing invariant.
By eliminating $\det U_{\rm MNS}^{*}$ from the inversion formula, it reduces to sixth order. 
While nine such invariants involving $\det U_{\rm MNS}$ can be constructed, this is the only one that induces a simple relation to the Dirac CP phase.

We also establish the connection between this formula and the well-known Jarlskog invariant \cite{Jarlskog:1985ht}. 
Dividing a given complex quantity by its absolute value, one obtains the phase of the quantity as,  
\begin{align}
e^{ i \d} =  \frac{ U_{e1} U_{e2} U_{\m 3} U_{\t 3} }{ U_{e3} \det U_{\rm MNS} } 
\abs{ \frac{ U_{e3} \det U_{\rm MNS} }{ U_{e1} U_{e2} U_{\m 3} U_{\t 3}} }
=  \frac{ U_{e1} U_{e2} U_{\m 3} U_{\t 3}  U_{e3}^{*} \det U_{\rm MNS}^{*}}{ |U_{e1} U_{e2} U_{\m 3} U_{\t 3}  U_{e3}| } 
 \, . 
\end{align}
Here we used the identity $ \det U_{\rm MNS}  \det U_{\rm MNS}^{*} = | \det U_{\rm MNS}|^{2} = 1$. 
To obtain the combination $U_{e1} U_{\t 3} U_{e 3}^{*} U_{\t 1}^{*}$, 
another element of the matrix inversion formula 
$U_{e 2} \det U_{\rm MNS}^{*} = U_{\m 3}^* U_{\t 1}^* - U_{\m 1}^* U_{\t 3}^* $
yields 
\begin{align}
e^{i \d} = 
\frac{ U_{e1}  U_{\m 3} U_{\t 3}  U_{e3}^{*} (U_{\m 3}^* U_{\t 1}^* - U_{\m 1}^* U_{\t 3}^*)}{ |U_{e1} U_{e2} U_{\m 3} U_{\t 3}  U_{e3}| } 
= \frac{| U_{\m 3}|^{2} U_{e1}  U_{\t 3}  U_{e3}^{*} U_{\t 1}^* - |U_{\t 3}|^{2} U_{e1}  U_{\m 3} U_{e3}^{*}  U_{\m 1}^*   }{ |U_{e1} U_{e2} U_{\m 3} U_{\t 3}  U_{e3}| }  \, . 
\end{align}
Taking the imaginary parts of both sides introduces the Jarlskog invariant $J \equiv \Im [U_{e1}  U_{\t 3}  U_{e3}^{*} U_{\t 1}^*] =  - \Im[U_{e1}  U_{\m 3} U_{e3}^{*}  U_{\m 1}^*  ] $.
From the identity $|U_{\m 3}|^{2} + |U_{\t 3}|^{2} = 1 - |U_{e 3}^{2}|$, it follows that 
\begin{align}
\Im e^{i\d} =  \frac{  (1 - |U_{e 3}^{2}| ) J}{ |U_{e1} U_{e2} U_{\m 3} U_{\t 3}  U_{e3}| } 
= \sin \d \, . 
\end{align}
The final equality follows from the standard expression $J = c_{12} s_{12} c_{23} s_{23} c_{13}^{2} s_{13} \sin \d$ with the observed mixing angles $s_{ij}, c_{ij}$. 
Therefore, the extracted $\sin \d$ is exactly the same as that obtained from the Jarlskog invariant.

As an application, we derive a general perturbative relation for the Dirac CP phase.
The lepton mixing matrix is defined as $U_{\rm MNS} \equiv U_e^\dagger U_\nu$, where $U_e$ and $U_\nu$ diagonalize the charged-lepton and neutrino mass matrices, respectively. 
These matrices can be parametrized as
$ U_{\nu,e} = \Phi_{\nu,e}^L \, U_{\nu,e}^0 \, \Phi_{\nu,e}^R, $ 
where $U_{\nu,e}^0$ are in the PDG standard form, and $\Phi_{\nu,e}^{L,R}$ are diagonal phase matrices.
By combining the left-handed phases as $\Phi^L \equiv \Phi_e^{L\dagger} \Phi_\nu^L$, the matrix becomes
\[
U_{\rm MNS} = \Phi_e^{R\dagger} \, U_e^{0\dagger} \, \Phi^L \, U_\nu^0 \, \Phi_\nu^R.
\]

In many grand unified scenarios, $U_e$ is expected to exhibit small mixings similar to the CKM matrix. %
Accordingly, we adopt the following approximation:

\textbf{Approximation:} The mixing angles in $U_e^0$ are treated perturbatively to first order.

\textbf{Justification:} 
If the charged-lepton Yukawa matrix $Y_e$ enjoys chiral symmetries acting on the first and second generations --- namely, if $Y_e = D_L Y_e D_R$ holds for $D_{L,R} = \mathrm{diag}(e^{i\phi_{L,R}^{1}}, e^{i\phi_{L,R}^{2}}, 1)$ --- then the corresponding mixing angles and singular values vanish exactly.
Although such symmetries are broken, as long as their breaking is small, the mixing angles are suppressed by powers of the small singular values and can be treated perturbatively.

To simplify the calculation, we redefine the unitary matrix of charged leptons as 
$U_{e}^{1 \dg} = \Phi^{L \dg} U_{e}^{0 \dg} \Phi^{L}$ without loss of generality.
Since $U_e^0$ has only small mixing angles, it is parameterized to first order as 
\begin{align}
U_{e}^{1} 
 \simeq 
\begin{pmatrix}
1 & s_{12}^{e} e^{- i \r_{12}} & s_{13}^{e}  e^{- i \r_{13}} \\
- s_{12}^{e}  e^{i \r_{12}}& 1 & s_{23}^{e}  e^{- i \r_{23}} \\
- s_{13}^{e}  e^{i \r_{13}} & - s_{23}^{e}  e^{i \r_{23}} & 1 \\
\end{pmatrix}  , 
\end{align}
where $s_{ij}^{f} \equiv \sin \th_{ij}^{f} \, , \, c_{ij}^{f} \equiv \cos \th_{ij}^{f}$, and $\r_{ij}$ are phases. 
The right-handed phases $\Phi_{\nu,e}^R$ is omitted because they do not contribute to the Dirac phase.

We perform a perturbative treatment in which the charged-lepton mixing angles $s_{ij}^{e}$ are sufficiently smaller compared to the neutrino mixing angles $s_{ij}^{\n}$.
Expanding arguments of the matrix elements of $U_{\rm MNS} = U_e^{1\dagger} U_\nu^0$ to first order in $s_{ij}^{e}$, we obtain:
\begin{align}
\arg U_{e1} & \simeq  s^e_{12}
\frac{ c^{\nu }_{12} s^{\nu }_{13} s^{\nu }_{23}\sin  \left(\delta _{\nu } - \rho _{12}\right) 
- c^{\nu }_{23}  s^{\nu }_{12} \sin \rho _{12} }{c^{\nu }_{12} c^{\nu }_{13}}
%%%
+s^e_{13} \frac{ c^{\nu }_{12} c^{\nu }_{23} s^{\nu }_{13} \sin  \left(\delta _{\nu }
- \rho _{13}\right) + s^{\nu }_{12} s^{\nu }_{23} \sin \rho _{13} }{c^{\nu }_{12} c^{\nu }_{13}} \, ,   \\
%%%%%%%%%%%%%%%%%%%%
\arg U_{e2} & \simeq
s^e_{12} \frac{ s^{\nu }_{12} s^{\nu }_{13} s^{\nu }_{23} \sin  (\delta _{\nu } - \rho _{12} ) 
+ c^{\nu }_{12} c^{\nu }_{23} \sin \rho _{12} }{c^{\nu }_{13} s^{\nu }_{12}}
%%%
+s^e_{13} \frac{ c^{\nu }_{23} s^{\nu }_{12} s^{\nu }_{13} \sin (\delta _{\nu } - \rho _{13} ) 
- c^{\nu }_{12}  s^{\nu }_{23}  \sin \rho _{13} }{c^{\nu }_{13} s^{\nu }_{12}} \, ,   \\
 %%%%%%%%%%%%%%%%%%%%
\arg U_{\m 3} & \simeq
s^e_{23} \frac{c^{\nu }_{23} \sin \rho _{23} }{s^{\nu }_{23}} 
-s^e_{12} \frac{ s^{\nu }_{13} \sin (\delta _{\nu } - \rho _{12} )}{c^{\nu }_{13} s^{\nu }_{23}} \, ,   
 %%%%%%%%%%%%%%%%%%%%
~ \arg U_{\t 3} \simeq 
s^e_{23} \frac{ s^{\nu}_{23} \sin \rho _{23} }{c^{\nu }_{23}}  
-s^e_{13} \frac{s^{\nu }_{13} \sin  (\delta _{\nu } - \rho _{13} )}{c^{\nu }_{13} c^{\nu }_{23}}  \, , \\
 %%%%%%%%%%%%%%%%%%%%
\arg U_{e3} & \simeq
-\delta _{\nu }
-s^e_{13} \frac{c^{\nu }_{13} c^{\nu }_{23}  \sin (\delta _{\nu } - \rho _{13} )}{s^{\nu }_{13}}
%%%
-s^e_{12} \frac{c^{\nu }_{13}  s^{\nu }_{23} \sin (\delta _{\nu } - \rho _{12} )}{s^{\nu }_{13}} \label{Ue3} \, . 
\end{align}
Since each contribution vanishes in the limit $s_{ij}^{e} \to 0$, the CP phases $\rho_{ij}$ appear at first  order in the corresponding $s_{ij}^{e}$.

In the present case, with $\arg (\det U_{\rm MNS}) = 0$, the Dirac phase $\delta$ is expanded as
\begin{align}
\d & \simeq \delta _{\nu } 
+ s^e_{12} D_{12} +s^e_{13} D_{13} + s^e_{23} \frac{  \sin \rho _{23}}{c^{\nu }_{23} s^{\nu }_{23}} \, , 
\end{align}
where the coefficients $D_{12}$ and $D_{13}$ are given by
\begin{align}
D_{12} & =  \frac{ (c^{\nu }_{13} )^2 (s^{\nu }_{23} )^2+\left(2 (s^{\nu }_{23} )^2-1\right) (s^{\nu }_{13} )^2  }{s^{\nu }_{13} s^{\nu }_{23} c^{\nu }_{13}} \sin (\delta _{\nu }- \rho _{12} )
%%%
+\frac{   (c^{\nu }_{12} )^2 - (s^{\nu }_{12} )^2  }{c^{\nu }_{12} s^{\nu }_{12} c^{\nu }_{13}} c^{\nu }_{23}  \sin \rho _{12}  \, , \\
%%%%%%%
D_{13} & =   \frac{ (c^{\nu }_{13} )^2 (c^{\nu }_{23} )^2 + \left(2 (c^{\nu }_{23} )^2-1 \right) (s^{\nu }_{13} )^2 }{s^{\nu }_{13}  c^{\nu }_{23} {c^{\nu }_{13}}} \sin (\delta _{\nu } - \rho _{13} )
%%%
- \frac{  (c^{\nu }_{12})^2 - (s^{\nu }_{12})^2  }{c^{\nu }_{12} s^{\nu }_{12} {c^{\nu }_{13}}} 
 s^{\nu }_{23}  \sin \rho _{13}  \, .
\end{align}
This expression is consistent with the result obtained by a perturbative expansion of the Jarlskog invariant \cite{Jarlskog:1985ht}.
A shift of Dirac phase $\epsilon = \delta - \delta_{\nu}$ is extracted using the addition formula
\begin{align}
\sin \d & = { \Im [U_{\m2} U_{\t 3} U_{\m 3}^{*} U_{\t 2}^{*} ] \over c_{12} s_{12} c_{23} s_{23} c_{13}^{2} s_{13} } \simeq \sin \d_{\n} + \e \cos  \d_{\n}  \, ,  \\
\To ~ \e & \simeq  s^e_{12} D_{12} +s^e_{13} D_{13} + s^e_{23} \frac{  \sin \rho _{23}}{c^{\nu }_{23} s^{\nu }_{23}} \, . 
\end{align}
Here, the observed mixing angles $c_{ij}, s_{ij}$ are expressed in terms of the mixing of leptons $c_{ij}^{\nu}, s_{ij}^{\nu}$ and $s_{ij}^{e}$.

We further show that each contribution from the charged leptons can reach 
$O(10^{\circ})$ or larger, and are therefore potentially observable by future experiments. 
Neglecting terms beyond first order in $s_{ij}^{e}$,  replacements $c_{ij}^{\nu} \to c_{ij}$ and $s_{ij}^{\nu} \to s_{ij}$ are valid. Using the observed values of the mixing angles~\cite{Esteban:2024eli}, we obtain
\begin{align}
\delta \simeq 
\delta _{\nu } + 2 s^e_{23}  \sin \rho _{23} 
+ s^e_{12}  [4.66 \sin (\delta _{\nu } - \rho _{12} ) + 0.51  \sin \rho _{12} ]
+ s^e_{13} [4.66 \sin (\delta _{\nu } - \rho _{13} ) - 0.51  \sin \rho _{13} ]
 \, . 
\end{align}
This expression allows us to assess the observational impact of each relative phase $\rho_{ij}$. For perturbative values $s_{ij}^{e} \lesssim 0.1$, the contributions from $\rho_{12}$ and $\rho_{13}$ remain below approximately 0.5, while that from $\rho_{23}$ is below 0.2.
In particular, if the charged-lepton mixing $U_e$ is close to the CKM matrix, contributions from $s_{13}^{e} \lesssim 0.01$ are at most 0.05 $\sim O(3^{\circ})$ and negligible. 
In this case, the previous analysis remains sufficiently general \cite{Yang:2024ulq}, and the total shift $\e$ induced by $s_{12}^{e}$ and $s_{23}^{e}$ is bounded by $\sim 0.7$ (or about $45^\circ$) at most.   

The dominant terms arise from $\arg U_{e3}$ in Eq.~(\ref{Ue3}), 
because $s_{13}^{\nu}$ enters its denominator and the factors $c_{23} / s_{13} \sim s_{23} / s_{13} \sim 5$ numerically enhance them. In addition, the term proportional to $s^e_{23} \sin \rho_{23} / (c_{23} s_{23})$, arising from $\arg U_{\mu 3} + \arg U_{\tau 3}$, also provides a non-negligible correction.

Finally, we comment on the validity of the perturbative treatment. Since this analysis assumes that the charged-lepton angles $s_{ij}^{e}$ are small compared to the neutrino angles $s_{ij}^{\nu}$,
the approximation becomes less accurate when some ratios $s^{e}_{kl} / s^{\n}_{ij}$ are $O(1)$. 
While $s_{13}^{\n}$ seems to be most sensitive to perturbative effects, the contribution from $s_{23}^{\n}$ is relatively less affected \cite{Yang:2024ulq}.  
Nevertheless, for variables where perturbation theory can break down, it suffices to return to Eq.~(\ref{d2}) and perform the expansion only in other parameters.

%%%%%%%%%%%%%%
\section{Conclusion}
%%%%%%%%%%%%%%

In this letter, we derive a rephasing invariant formula 
%$\d  = \arg ( U_{e1} U_{e2} U_{\m 3} U_{\t 3} U_{e3}^{*} \det U_{\rm MNS}^{*} )$ 
to extract the Dirac CP phase $\d$ directly from the leptonic mixing matrix $U_{\rm MNS}$ in an arbitrary basis of phases. 
This new type of invariant enables a more direct analysis of  the CP phase and offers wider theoretical exploration of CP violation. 
Furthermore, performing a general perturbative expansion in terms of the small mixing angles $s_{ij}^{e}$ of charged leptons, 
we present a compact relation between the phase $\delta$ and the underlying neutrino CP phase $\delta_{\nu}$ with $s_{ij}^{e}$.
%$\delta = \delta_{\nu} + s_{12}^{e} D_{12} + s_{13}^{e} D_{13} + s_{23}^{e} D_{23}$. 
%
For small mixing angles $s_{ij}^{e} \lesssim 0.1$, corrections in $\delta$ from $s_{12}^{e}, s_{13}^{e}$, and $s_{23}^{e}$ can reach up to $0.5 \sim 30^\circ$, $0.5 \sim 30^\circ$, and $0.2 \sim 10^\circ$, respectively. Such corrections lie within the expected sensitivity of upcoming experiments
and therefore the impact of charged-lepton phases should be properly taken into account. 
The presented formulation provides a model-independent framework to interpret CP-violating phases of quarks and leptons, and it will be a valuable tool for studies of flavor physics and unified theories.

%%%%%%%%%%%%%%
\section*{Acknowledgment} 
%%%%%%%%%%%%%%
The study is partly supported by the MEXT Leading Initiative for Excellent Young Researchers Grant Number JP2023L0013.

%\bibliographystyle{bib/h-physrev50}
%\bibliography{bib/fourzero,bib/onezero,bib/refsym,bib/mutausym,bib/PSGUT,bib/StrongCP,bib/LR,bib/GCP,bib/U(2),bib/flaxion,bib/minimal-natural,bib/chiral, bib/T2HK,bib/CKM2MNS}

\end{document}